\input{aipcheck}

\documentclass[
    ,final            % use final for the camera ready runs
  ]
  {aipproc}

\usepackage{graphicx,epsfig,color}

\newcommand{\bc}{\begin{center}}
\newcommand{\ec}{\end{center}}
\newcommand{\be}{\begin{equation}}
\newcommand{\ee}{\end{equation}}
\newcommand{\bea}{\begin{eqnarray}}
\newcommand{\eea}{\end{eqnarray}}
\newcommand{\ba}{\begin{array}}
\newcommand{\ea}{\end{array}}
\newcommand{\lb}{\label}
\newcommand{\rf}{\ref}
\newcommand{\bfg}{\begin{figure}[htbp]}
\newcommand{\efg}{\end{figure}}

\layoutstyle{6x9}

\begin{document}

\title{The gauge invariant quark Green's function \protect \\
in two-dimensional QCD}


\author{H. Sazdjian}{
  address={Institut de Physique Nucl\'eaire, CNRS/IN2P3,\\
Universit\'e Paris-Sud 11, F-91405 Orsay, France\\
E-mail: sazdjian@ipno.in2p3.fr}
}

\begin{abstract}
The gauge invariant quark Green's function, defined with a 
path-ordered phase factor along a straight-line, is studied in 
two-dimensional QCD in the large-$N_c$ limit by means of an exact 
integro\-differential equation. It is found to be infrared finite 
with singularities represented by an infinite number of threshold 
type branch points with a power of -3/2, starting at positive mass 
squared values. The Green's function is analytically determined.
\end{abstract}

\maketitle

%%%%%%%%%%%%%%%%%%%%%%%%%%%%%%%%%%%%%%%%%%%%
%% MAINMATTER
%%%%%%%%%%%%%%%%%%%%%%%%%%%%%%%%%%%%%%%%%%%%

\section{Gauge invariant Green's functions}

Gauge invariant Green's functions are expected to provide more
reliable information about the physical properties of observables
than gauge variant ones. In QCD, they are defined with the aid
of path-ordered gluon field phase factors \cite{m,nm}. We report 
in this talk on recent results obtained for the gauge invariant
quark Green's function in two-dimensional QCD \cite{s1,s2}. 
Using for the paths skew-polygonal lines, it is possible to derive 
for the latter in any dimensions an exact integrodifferential 
equation in functional form; its restriction to two dimensions
in the large $N_c$ limit then allows us to solve analytically the
above equation and to have an explicit check of the spectral
properties of the quark fields.
\par
For quarks, the gauge invariant two-point Green's function is defined 
as
\be \lb{e1}
S_{\alpha\beta}(x,x';C_{x'x})=-\frac{1}{N_c}\,\langle\overline 
\psi_{\beta}(x')\,U(C_{x'x};x',x)\,\psi_{\alpha}(x)\rangle,
\ee
$\alpha$ and $\beta$ being the Dirac spinor indices, while the 
color indices are implicitly summed; $U$ is a path-ordered gluon
field phase factor along a line $C_{x'x}$ joining a point $x$ to a 
point $x'$, with an orientation defined from $x$ to $x'$:
\be \lb{e2}
U(C_{x'x};x',x)=Pe^{{\displaystyle -ig\int_x^{x'} 
dz^{\mu}A_{\mu}(z)}}.
\ee
\par
Green's functions with paths along skew-polygonal lines are of
particular interest, since they can be decomposed into the
succession of simpler straight line segments. For such lines 
with $n$ sides and $n-1$ junction points $y_1$, $y_2$, $\ldots$, 
$y_{n-1}$ between the segments, we define:
\be \lb{e3}
S_{(n)}(x,x';y_{n-1},\dots,y_1)=-\frac{1}{N_c}\,\langle\overline \psi(x')
U(x',y_{n-1})U(y_{n-1},y_{n-2})\ldots U(y_1,x)\psi(x)\rangle,
\ee
where now each $U$ is along a straight line segment.
The simplest such function corresponds to $n=1$, for which the points
$x$ and $x'$ are joined by a single straight line:
\be \lb{e4}
S_{(1)}(x,x')\equiv S(x,x')=-\frac{1}{N_c}\,\langle\overline \psi(x')
\,U(x',x)\,\psi(x)\rangle.
\ee
(We shall generally omit the index 1 from that function.)
\par
 
\section{Integrodifferential equation}
 
To quantize the theory one may proceed in two steps. First, one
integrates with respect to the quark fields. This produces in
various terms the quark propagator in the presence of the gluon 
field. Then one integrates with respect to the gluon field through
Wilson loops \cite{w,p,mm1,mm2,mgd,mk}. To achieve the latter
program, we use for the quark propagator in external field a
representation which involves phase factors along straight lines
together with the full gauge invariant quark Green's function
\cite{s1,js}. The latter feature allows implicit summation of 
self-energy effects at each step of the operation. This representation
is a generalization of the one introduced by Eichten and Feinberg
when dealing with the heavy quark limit \cite{ef}. 
\par
The quark propagator in the external gluon field is expanded
around the following gauge covariant quantity:
\be \lb{e5}  
\Big[\widetilde S(x,x')\Big]_{\ b}^a\equiv 
S(x,x')\Big[U(x,x')\Big]_{\ b}^a.
\ee
It is possible to set up an integral equation realizing iteratively
the previous expansion. Its systematic use leads to the derivation of
functional relations between the Green's functions $S_{(n)}$ 
(skew-polygonal line with $n$ segments) and $S$ (one segment).
\par
Using then the equations of motion relative to the Green's functions, 
one establishes the following equation for $S(x,x')$ \cite{s1}:
\bea \lb{e6}
& &(i\gamma.\partial_{(x)}-m)S(x,x')=i\delta^4(x-x')
+i\gamma^{\mu}\Big\{K_{2\mu}(x',x,y_1)\,S_{(2)}(y_1,x';x)\nonumber \\
& &\ +\sum_{n=3}^{\infty}K_{n\mu}(x',x,y_1,\ldots,y_{n-1})\,
S_{(n)}(y_{n-1},x';x,y_1,\ldots,y_{n-2})\Big\},
\eea
where the kernel $K_n$ ($n=2,3,\ldots$) contains globally $n$ 
derivatives of Wilson loop averages with skew-polygonal contours and 
also the Green's function $S$ and its derivative. The Green's functions 
$S_{(n)}$ being themselves related to the simplest Green's function $S$ 
through series expansions resulting from functional relations, 
Eq. (\rf{e6}) is ultimately an integrodifferential equation for $S$. 
One expects that the kernels with small number of derivatives will 
provide the most salient contributions. Therefore, the first kernel $K_2$ 
in Eq. (\rf{e6}) would contain the leading effect of the interaction.  
\par
 
\section{Interest of the quark Green's function}

The interest of the gauge invariant quark Green's function is
related to its particular status. If the theory is confining, 
then it is not possible to cut the Green's function and to saturate 
it with a complete set of physical states (hadrons), which are color 
singlets. Intermediate states are necessarily colored states.
This would suggest that the Green's function does not have any
singularity. However, the equation it satisfies [Eq. (\rf{e6})],
derived from the QCD Lagrangian, contains singularities, generated
by the free quark propagator (the inverse of the Dirac operator
in the left-hand side of Eq. (\rf{e6})).
\par
The above paradoxical situation is overcome with the acceptance 
that quarks and gluons continue forming a complete set of states
with positive energies and could be used for any saturation
scheme of intermediate states. It is the resolution of the 
equations of motion which should indicate to us how the related 
singularities combine to form the complete solutions.
\par
Therefore, the knowledge of the gauge invariant quark Green's
function provides a direct information about the effect of 
confinement in the colored sector of quarks.
\par
 
\section{Spectral functions}

Green's functions with paths along straight lines are dependent
only on the end points of the paths. The transition is then simple
to momentum space by Fourier transformation.
\par
It is advantageous to consider for that purpose the path-ordered 
phase factor in its representation given by the formal series expansion 
in terms of the coupling constant $g$.
\par
Using for each term of the series, together with the quark fields,
the spectral analysis with intermediate states and causality, one
arrives at a generalized form of the K\"all\'en--Lehmann
representation for the Green's function $S$ in momentum space, in
which the cut lies on the positive real axis staring from the quark 
mass squared $m^2$ and extending to infinity \cite{k,l,wght,schwb,thftv}.
\par
Taking into account translation invariance, we introduce the
Fourier transform of the Green's function $S$ into momentum space:
\be \lb{e7}
S(x,x')=S(x-x')=\int \frac{d^4p}{(2\pi)^4}\,
e^{{\displaystyle -ip.(x-x')}}\,S(p).
\ee 
$S(p)$ has the following representation in terms of real spectral 
functions $\rho_1^{(n)}$ and $\rho_0^{(n)}$ ($n=1,\ldots,\infty$):
\be \lb{e8}
S(p)=i\int_0^{\infty}ds'\,\sum_{n=1}^{\infty}\,
\frac{\big[\,\gamma.p\,\rho_1^{(n)}(s')+\rho_0^{(n)}(s')\,\big]}
{(p^2-s'+i\varepsilon)^n}.
\ee
Depending on the degrees of the singularities at threshold, 
simplifications may occur by integrations by parts, or otherwise
by summation, reducing the series into more compact forms.
\par

\section{Two-dimensional QCD}

Many simplifications occur in two-dimensional QCD at large $N_c$
\cite{thft1,thft2,ccg}. This theory is expected to have the
essential features of confinement observed in four dimensions,
with the additional simplification that asymptotic freedom is
realized here in a trivial way, since the theory is
super\-renormalizable. For simple contours, Wilson loop averages
in two dimensions are exponential functionals of the areas enclosed
by the contours \cite{kzkk,kzk,b}. Furthermore, at large $N_c$,
crossed diagrams and quark loop contributions disappear.
\par
It turns out that in two dimensions and at large $N_c$, only the 
lowest-order kernel $K_2$ survives in Eq. (\rf{e6}). The equation
reduces then to the following form \cite{s2}:
\bea \lb{e9}
& &(i\gamma.\partial-m)S(x)=i\delta^2(x)
-\sigma\gamma^{\mu}(g_{\mu\alpha}g_{\nu\beta}-g_{\mu\beta}g_{\nu\alpha}) 
x^{\nu}x^{\beta}\nonumber \\
& &\ \ \ \ \ \times\left[\,\int_0^1d\lambda\,\lambda^2\,S((1-\lambda)x)
\gamma^{\alpha}S(\lambda x)
+\int_1^{\infty}d\xi\,S((1-\xi)x)\gamma^{\alpha}S(\xi x)\,\right], 
\eea
where $\sigma$ is the string tension.
\par
The equation is solved by decomposing $S$ into Lorentz invariant
parts:
\be \lb{e10}
S(p)=\gamma.pF_1(p^2)+F_0(p^2),
\ee
or, in $x$-space:
\be \lb{e11}
S(x)=\frac{1}{2\pi}\Big(\frac{i\gamma.x}{r}\widetilde F_1(r)
+\widetilde F_0(r)\Big),\ \ \ \ \ r=\sqrt{-x^2}. 
\ee
\par
One obtains, with the introduction of the Lorentz invariant functions,
two coupled equations. Their resolution proceeds through several steps, 
mainly based on the analyticity properties resulting from the
spectral representation (\rf{e8}). The solutions are obtained in explicit
form for any value of the quark mass $m$.
\par
The covariant functions $F_1(p^2)$ and $F_0(p^2)$ are, for complex $p^2$:
\bea
\lb{e12}
& &F_1(p^2)=-i\frac{\pi}{2\sigma}\,\sum_{n=1}^{\infty}\,
b_n\,\frac{1}{(M_n^2-p^2)^{3/2}},\\
\lb{e13}
& &F_0(p^2)=i\frac{\pi}{2\sigma}\,\sum_{n=1}^{\infty}\,
(-1)^nb_n\,\frac{M_n}{(M_n^2-p^2)^{3/2}}.
\eea
The masses $M_n$ ($n=1,2,\ldots$) have positive values greater than
the quark mass $m$ and are labelled with increasing values with respect 
to $n$; their squares represent the locations of branch point 
singularities with power $-3/2$. The masses $M_n$ and the coefficients
$b_n$ satisfy an infinite set of coupled algebraic equations that are
solved numerically. Their asymptotic behaviors for large $n$, such that
$\sigma\pi n\gg m^2$, are:
\be \lb{e14}
M_n^2\simeq \sigma\pi n,\ \ \ \ \ \ \ 
b_n\simeq \frac{\sigma^2}{M_n}.
\ee
\par
In $x$-space, the solutions are:
\be \lb{e15}
\widetilde F_1(r)=\frac{\pi}{2\sigma}\,\sum_{n=1}^{\infty}\,
b_n\,e^{-M_nr},\ \ \ \ \
\widetilde F_0(r)=\frac{\pi}{2\sigma}\,\sum_{n=1}^{\infty}\,
(-1)^{n+1}b_n\,e^{-M_nr}.
\ee
[$r=\sqrt{-x^2}$.] 
\par
At high energies, the solutions satisfy asymptotic freedom \cite{pltz}:
\bea \lb{e16}
& &F_1(p^2)_{\stackrel{{\displaystyle =}}{p^2\rightarrow -\infty}}
\ \frac{i}{p^2}, \\
\lb{e17}
& &F_0(p^2)_{\stackrel{{\displaystyle =}}{p^2\rightarrow -\infty}}
\ \frac{im}{p^2},\ \ \ \ \ m\neq 0, \\
\lb{e18}
& &F_0(p^2)_{\stackrel{{\displaystyle =}}{p^2\rightarrow -\infty}}
\frac{2i\sigma}{N_c}\frac{\langle\overline\psi\psi\rangle}
{(p^2)^2},\ \ \ \ \ \ m=0,
\eea
where in the last equation we have introduced the one-flavor quark 
condensate.
\par  
We present in Fig. \rf{f1} the function $iF_0$ for spacelike $p$ and 
in Fig. \rf{f2} its real part for timelike $p$, for the case $m=0$. 
\par
%\vspace{0.25 cm}
\bfg
%\bc
\epsfig{file=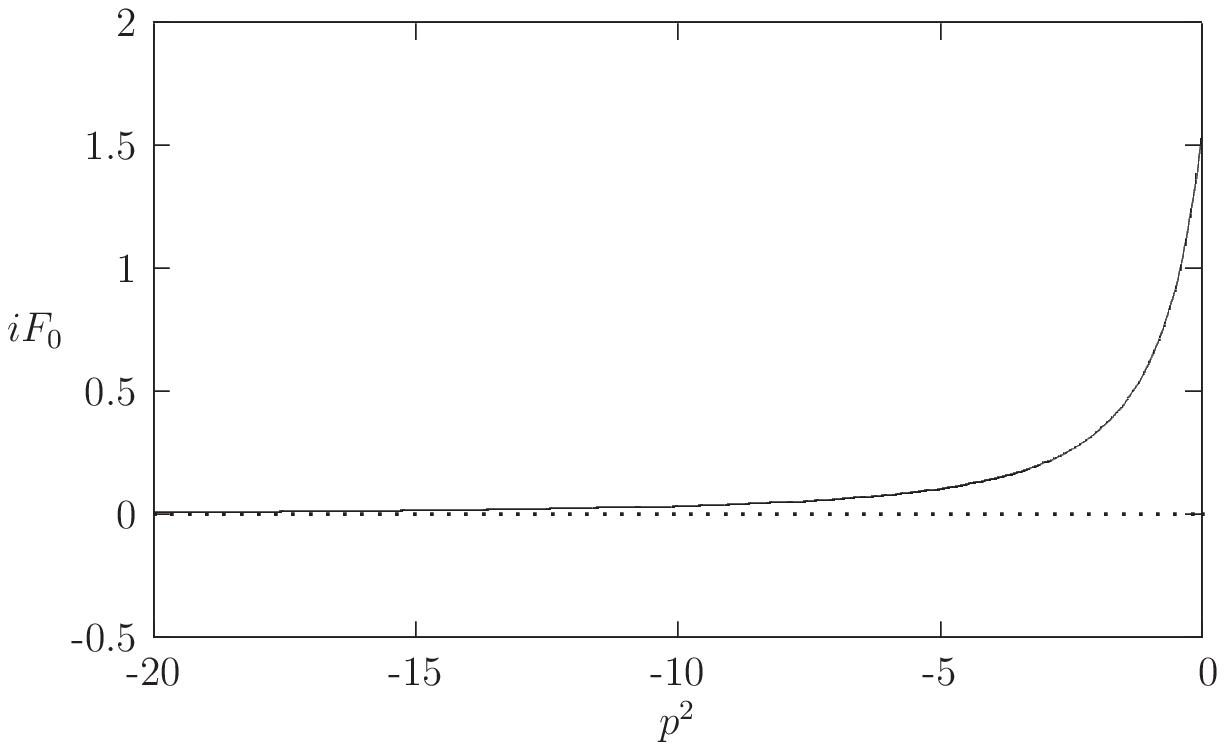,bbllx=80,bblly=470,bburx=530,bbury=720}
\caption{The function $iF_0$ for spacelike $p$, in mass unit of 
$\sqrt{\sigma/\pi}$, for $m=0$.} 
\lb{f1}
%\ec
\efg
\par
\bfg
%\bc
\epsfig{file=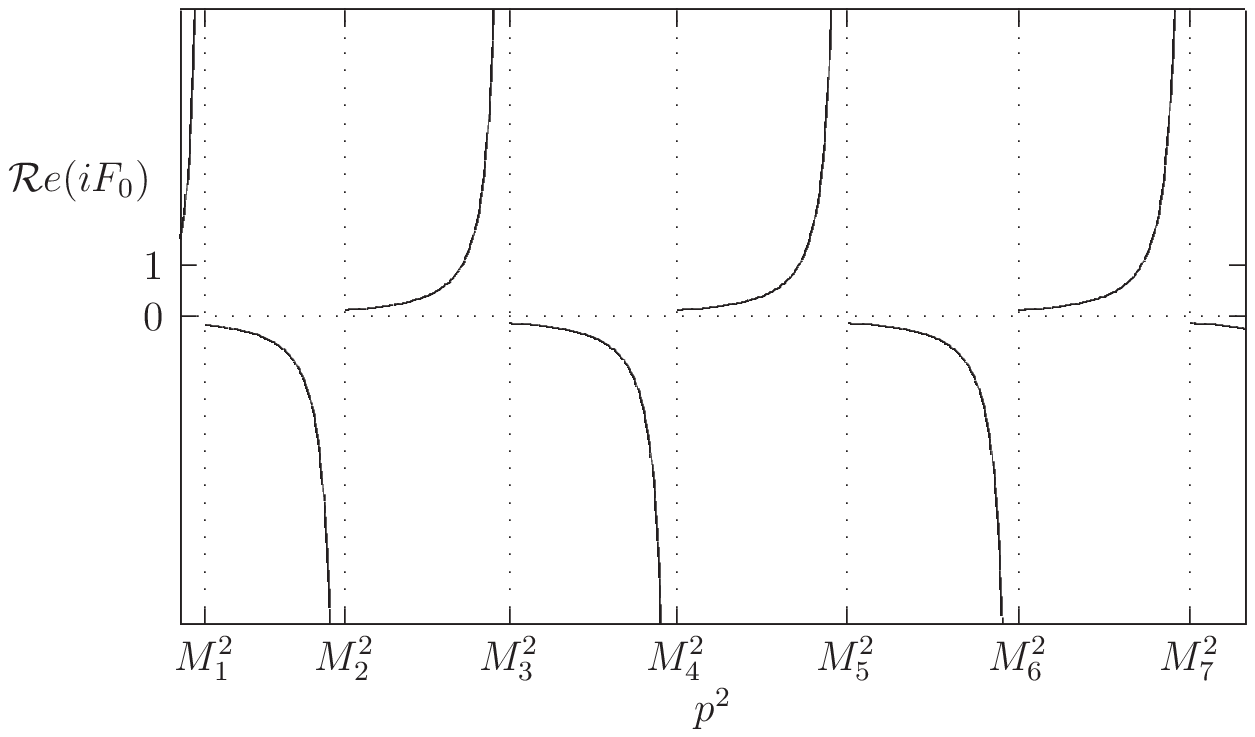,bbllx=80,bblly=470,bburx=530,bbury=720}
\caption{The real part of the function $iF_0$ for timelike $p$, in mass 
unit of $\sqrt{\sigma/\pi}$, for $m=0$.} 
\lb{f2}
%\ec
\efg
\par

\section{Conclusion}

The spectral functions of the quark Green's function are infrared
finite and lie on the positive real axis of $p^2$. No singularities 
in the complex plane or on the negative real axis have been found.
This means that quarks contribute like physical particles with positive 
energies. (In two dimensions there are no physical gluons.)
\par 
The singularities of the Green's function are represented by an
infinite number of threshold type singularities, characterized by 
a power of $-3/2$ and positive masses $M_n$ ($n=1,2,\ldots$). The
corresponding singularities are stronger than simple poles and this 
feature might mean difficulty in the observability of quarks as
asymptotic states.
\par
The threshold masses $M_n$ represent dynamically generated masses and 
maintain the scalar part of the Green's function at a nonzero value
even when the quark mass is zero.
\par

%%%%%%%%%%%%%%%%%%%%%%%%%%%%%%%%%%%%%%%%%%%%%%%%
%% BACKMATTER
%%%%%%%%%%%%%%%%%%%%%%%%%%%%%%%%%%%%%%%%%%%%%%%%

\begin{theacknowledgments}
This work was supported in part by the EU network FLAVIANET, under 
Contract No. MRTN-CT-2006-035482, and by the European Community 
Research Infrastructure Integrating Activity ``Study of Strongly 
Interacting Matter'' (acronym HadronPhysics2, Grant Agreement 
No. 227431), under the Seventh Framework Programme of EU.
I thank H. M. Fried and A. Pineda for stimulating discussions during
the Workshop.
\end{theacknowledgments}



\bibliographystyle{aipproc}   % if natbib is available

\bibliography{giqgf2d}

\IfFileExists{\jobname.bbl}{}
 {\typeout{}
  \typeout{******************************************}
  \typeout{** Please run "bibtex \jobname" to optain}
  \typeout{** the bibliography and then re-run LaTeX}
  \typeout{** twice to fix the references!}
  \typeout{******************************************}
  \typeout{}
 }

\end{document}